\documentclass[twocolumn,showpacs,aps,prl,superscriptaddress]{revtex4b5}

\usepackage{graphicx}
\usepackage{dcolumn}
\usepackage{amsmath}
\usepackage{epsfig}

\RequirePackage{xspace}

\usepackage{relsize}
\def\babar{\mbox{\slshape B\kern-0.1em{\smaller A}\kern-0.1em
    B\kern-0.1em{\smaller A\kern-0.2em R}}}

\def\Kbar  {\kern 0.2em\overline{\kern -0.2em K}{}\xspace}

\def\Kzb   {\ensuremath{\Kbar^0}\xspace}
\def\KzKzb {\ensuremath{K^0 \kern -0.16em \Kzb}\xspace}

\def\Dbar  {\kern 0.2em\overline{\kern -0.2em D}{}\xspace}

\def\Dzb   {\ensuremath{\Dbar^0}\xspace}
\def\DzDzb {\ensuremath{D^0 {\kern -0.16em \Dzb}}\xspace}

\def\Bbar  {\kern 0.18em\overline{\kern -0.18em B}{}\xspace}

\def\Bzb   {\ensuremath{\Bbar^0}\xspace}

\def\BzBzb {\ensuremath{B^0 {\kern -0.16em \Bzb}}\xspace}

\mathchardef\Upsilon="7107
\def\Y#1S{\ensuremath{\Upsilon{(#1S)}}\xspace}

\mathchardef\Deltares="7101
\mathchardef\Xi="7104
\mathchardef\Lambda="7103
\mathchardef\Sigma="7106
\mathchardef\Omega="710A
\def\Deltabar   {\kern 0.25em\overline{\kern -0.25em \Deltares}{}\xspace}
\def\Lbar {\kern 0.2em\overline{\kern -0.2em\Lambda\kern 0.05em}\kern-0.05em{}\xspace}
\def\Sigbar{\kern 0.2em\overline{\kern -0.2em \Sigma}{}\xspace}
\def\Xibar{\kern 0.2em\overline{\kern -0.2em \Xi}{}\xspace}
\def\Obar{\kern 0.2em\overline{\kern -0.2em \Omega}{}\xspace}
\def\Nbar{\kern 0.2em\overline{\kern -0.2em N}{}\xspace}
\def\Xb{\kern 0.2em\overline{\kern -0.2em X}{}}

\newcommand{\tev}{\ensuremath{\mathrm{\,Te\kern -0.1em V}}\xspace}
\newcommand{\gev}{\ensuremath{\mathrm{\,Ge\kern -0.1em V}}\xspace}
\newcommand{\mev}{\ensuremath{\mathrm{\,Me\kern -0.1em V}}\xspace}
\newcommand{\kev}{\ensuremath{\mathrm{\,ke\kern -0.1em V}}\xspace}
\newcommand{\ev}{\ensuremath{\mathrm{\,e\kern -0.1em V}}\xspace}
\newcommand{\gevc}{\ensuremath{{\mathrm{\,Ge\kern -0.1em V\!/}c}}\xspace}
\newcommand{\mevc}{\ensuremath{{\mathrm{\,Me\kern -0.1em V\!/}c}}\xspace}
\newcommand{\gevcc}{\ensuremath{{\mathrm{\,Ge\kern -0.1em V\!/}c^2}}\xspace}
\newcommand{\mevcc}{\ensuremath{{\mathrm{\,Me\kern -0.1em V\!/}c^2}}\xspace}

\def\mus  {\ensuremath{\rm \,\mus}\xspace}

\def\mus        {\ensuremath{\,\mu{\rm s}}\xspace}    
\def\gsim{{~\raise.15em\hbox{$>$}\kern-.85em
          \lower.35em\hbox{$\sim$}~}\xspace}
\def\lsim{{~\raise.15em\hbox{$<$}\kern-.85em
          \lower.35em\hbox{$\sim$}~}\xspace}

\def\pep2{PEP-II}

\xspace

\def\jetset74   {\mbox{\tt Jetset \hspace{-0.5em}7.\hspace{-0.2em}4}}

\newcommand{\BABARPubYear}    {01}
\newcommand{\BABARPubNumber}  {11}

\newcommand{\SLACPubNumber} {8823}
\newcommand{\LANLNumber} {010xxxx}

\def\figurebox#1#2#3{%
    \def\arg{#3}%
    \ifx\arg\empty
    {\hfill\vbox{\hsize#2\hrule\hbox to #2{\vrule\hfill\vbox to #1{\hsize#2\vfill}\vrule}\hrule}\hfill}%
    \else
    {\hfill\epsfbox{#3}\hfill}%
    \fi}

\long\def\inst#1{\par\nobreak\kern 4pt\nobreak
    {\it #1}\par\vskip 10pt plus 3pt minus 3pt}

\begin{document}

\preprint{\babar-PUB-\BABARPubYear/\BABARPubNumber} 
\preprint{SLAC-PUB-\SLACPubNumber} 

\begin{flushleft}
\babar-PUB-\BABARPubYear/\BABARPubNumber\\
SLAC-PUB-\SLACPubNumber\\
hep-ex/\LANLNumber\\[20mm]
\end{flushleft}

\title{
\vskip 10mm
{\Large \bf
Measurement of the decays \boldmath{$B\rightarrow\phi K$} 
and \boldmath{$B\rightarrow\phi K^*$}
} 
\begin{center} 
\vskip 10mm
The \babar\ Collaboration
\end{center}
}


%
\author{B.~Aubert}
\author{D.~Boutigny}
\author{J.-M.~Gaillard}
\author{A.~Hicheur}
\author{Y.~Karyotakis}
\author{J.P.~Lees}
\author{P.~Robbe}
\author{V.~Tisserand}
\affiliation{Laboratoire de Physique des Particules, F-74941 Annecy-le-Vieux, France }
\author{A.~Palano}
\affiliation{Universit\`a di Bari, Dipartimento di Fisica and INFN, I-70126 Bari, Italy }
\author{G.P.~Chen}
\author{J.C.~Chen}
\author{N.D.~Qi}
\author{G.~Rong}
\author{P.~Wang}
\author{Y.S.~Zhu}
\affiliation{Institute of High Energy Physics, Beijing 100039, China }
% ...

\date{March 29, 2001}

\begin{abstract}
We have observed the decays $B\rightarrow\phi K$ and $\phi K^*$
in a sample of over 45 million $B$ mesons collected with the 
\babar\ detector at the PEP-II collider.
The measured branching fractions are
${\cal B} (B^+ \rightarrow \phi K^+) = (7.7^{+1.6}_{-1.4}\pm 0.8)\times 10^{-6}$,
${\cal B} (B^0 \rightarrow \phi K^0) = (8.1^{+3.1}_{-2.5}\pm 0.8)\times 10^{-6}$,
${\cal B} (B^+ \rightarrow \phi K^{*+}) = (9.7^{+4.2}_{-3.4}\pm 1.7)\times 10^{-6}$, and
${\cal B} (B^0 \rightarrow \phi K^{*0}) = (8.6^{+2.8}_{-2.4}\pm 1.1)\times 10^{-6}$.
We also report the upper limit 
${\cal B} (B^+ \rightarrow \phi\pi^+) < 1.4\times 10^{-6}$ (90\% CL).
\end{abstract}
\pacs{ 
13.25.Hw, 
13.25.-k, 
14.40.Nd  
}


\maketitle

\par
\begin{figure}
\begin{center}
\centerline{\epsfig{figure=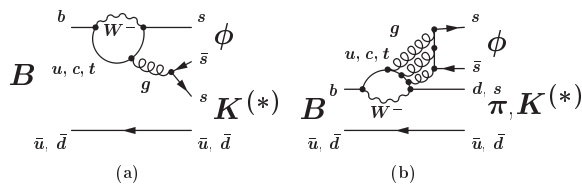,height=2.5cm}}
\caption{
Gluonic penguin diagrams describing the decays 
$B \rightarrow \phi K$, $\phi K^*$, and $\phi\pi$: 
 (a) internal (b) flavor-singlet.}
\label{fig:diagram} 
\end{center}
\end{figure}

The decays of $B$ mesons into charmless hadronic final states provide
important information for the study of $CP$ violation and the search for
new physics.
Decays into final states
containing a $\phi$ meson are particularly interesting because they
are dominated by $b\rightarrow s(d)\bar{s}s$ penguins
(Fig.~\ref{fig:diagram}), with gluonic and electroweak contributions,
while other Standard Model contributions are highly suppressed \cite{theory1}.
These modes thus provide a direct measurement of the penguin process,
with potential benefits to estimates of direct $CP$ violation.
They also allow an independent measurement of $\sin 2\beta$ \cite{theory3}.
Comparison of the value of $\sin 2 \beta$ obtained from these 
modes with that from charmonium modes, 
as well as various tests of isospin relationships, can probe for 
new physics \cite{theory4,theory5}. 

In this paper we present measurements of four
such decays: $B^+\rightarrow\phi K^+$, $B^0\rightarrow\phi K^0$,
$B^+\rightarrow\phi K^{*+}$, and $B^0\rightarrow\phi K^{*0}$.
Charge conjugate states are assumed throughout this paper
and measured branching fractions are averaged accordingly.
The $\phi K^+$ and $\phi K^{*0}$ modes have been previously seen~\cite{firstphik}.

The data were collected with the \babar\ detector~\cite{babar}
at the PEP-II asymmetric $e^+e^-$ collider~\cite{pep}
located at the Stanford Linear Accelerator Center.
The results presented in this paper are based on data taken
in the 1999--2000 run. An integrated
luminosity of 20.7~fb$^{-1}$ was recorded corresponding to 
22.7 million $B\overline{B}$ pairs at the $\Upsilon (4S)$ resonance
(``on-resonance'') and 2.6~fb$^{-1}$ about 40~MeV below
this energy (``off-resonance''). 

The asymmetric beam configuration in the laboratory frame
provides a boost to the $\Upsilon(4S)$
increasing the momentum range of the $B$-meson decay products
up to 4.3~GeV/$c$.
Charged particles are detected and their momenta are measured
by a combination of a silicon vertex tracker (SVT) consisting 
of five double-sided layers and a 40-layer central drift chamber 
(DCH), both operating in a 1.5~T solenoidal magnetic field. 
With the SVT, a position resolution of about 40~$\mu$m is 
achieved for the highest momentum charged particles near the 
interaction point, allowing the precise determination of 
decay vertices.
The tracking system covers 92\% of the solid angle
in the center-of-mass system (CM).
The track finding efficiency is, on average, (98$\pm$1)\% for momenta
above 0.2~GeV/$c$ and polar angle greater than 0.5~rad. 
Photons are detected by a CsI electromagnetic calorimeter (EMC), which
provides excellent angular and energy resolution with high efficiency for 
energies above 20~MeV~\cite{babar}.

Charged particle identification is provided by the average 
energy loss ($dE/dx$) in the tracking devices  and
by a unique, internally reflecting ring imaging 
Cherenkov detector (DIRC) covering the central region. 
A Cherenkov angle $K$--$\pi$ separation of better than 4$\sigma$ is 
achieved for tracks below 3~GeV/$c$ momentum, decreasing to 
2.5$\sigma$ at the highest momenta in our final states. 
Electrons are identified with the use of the EMC.

Hadronic events are selected based on track multiplicity and 
event topology. We fully reconstruct $B$ meson 
candidates from their charged and neutral 
decay products, where we recover the intermediate states
$\pi^0\rightarrow \gamma\gamma$,
$K^0\rightarrow K^0_S\rightarrow\pi^+\pi^-$, 
$\phi\rightarrow K^+K^-$,
$K^{*+}\rightarrow K^0\pi^+$ or $K^+\pi^0$, and 
$K^{*0}\rightarrow K^+\pi^-$.
Candidate charged tracks are required to originate 
from the interaction point (within 
10~cm along the beam direction and 1.5~cm in the transverse plane),
and to have at least 12 DCH hits 
and a minimum transverse momentum of 0.1~GeV/$c$. 
Looser criteria are applied to tracks forming $K^0_S$ candidates
to allow for displaced decay vertices.
Kaon tracks are distinguished from pion and proton tracks via a
likelihood ratio that includes, for momenta below 0.7~GeV/$c$, 
$dE/dx$ information from the SVT and DCH, and, for higher
momenta, the Cherenkov angle and number of photons
as measured by the DIRC. 
A kaon (pion) candidate is any track not identified as a proton or 
pion (kaon).

We reconstruct $\pi^0$ mesons as pairs of photons
with a minimum energy deposition of 30~MeV. 
The typical width of the reconstructed $\pi^0$ mass is 7~MeV/$c^2$.
A $\pm$15~MeV/$c^2$ interval is applied to select $\pi^0$ candidates. 

We combine pairs of tracks with opposite charge from a common
vertex to form $K^0_S$, $\phi$, and $K^{*0}$ candidates.
The selection of $K^0_S$ candidates is based on the invariant two-pion mass 
($|M_{\pi\pi} - m_{K^0}|<$ 10~MeV/$c^2$), 
the angle $\alpha$ between the reconstructed flight 
and momentum directions in the plane transverse to the beam direction
($\cos\alpha >$ 0.999), 
and the measured lifetime significance ($\tau/\sigma_\tau >$ 3).
For the softer $K^0_S$ from $K^{*+}$ decays we relax the criteria
to 12~MeV/$c^2$ and $\cos\alpha >$ 0.995.

\begin{figure}
\centerline{\epsfig{figure=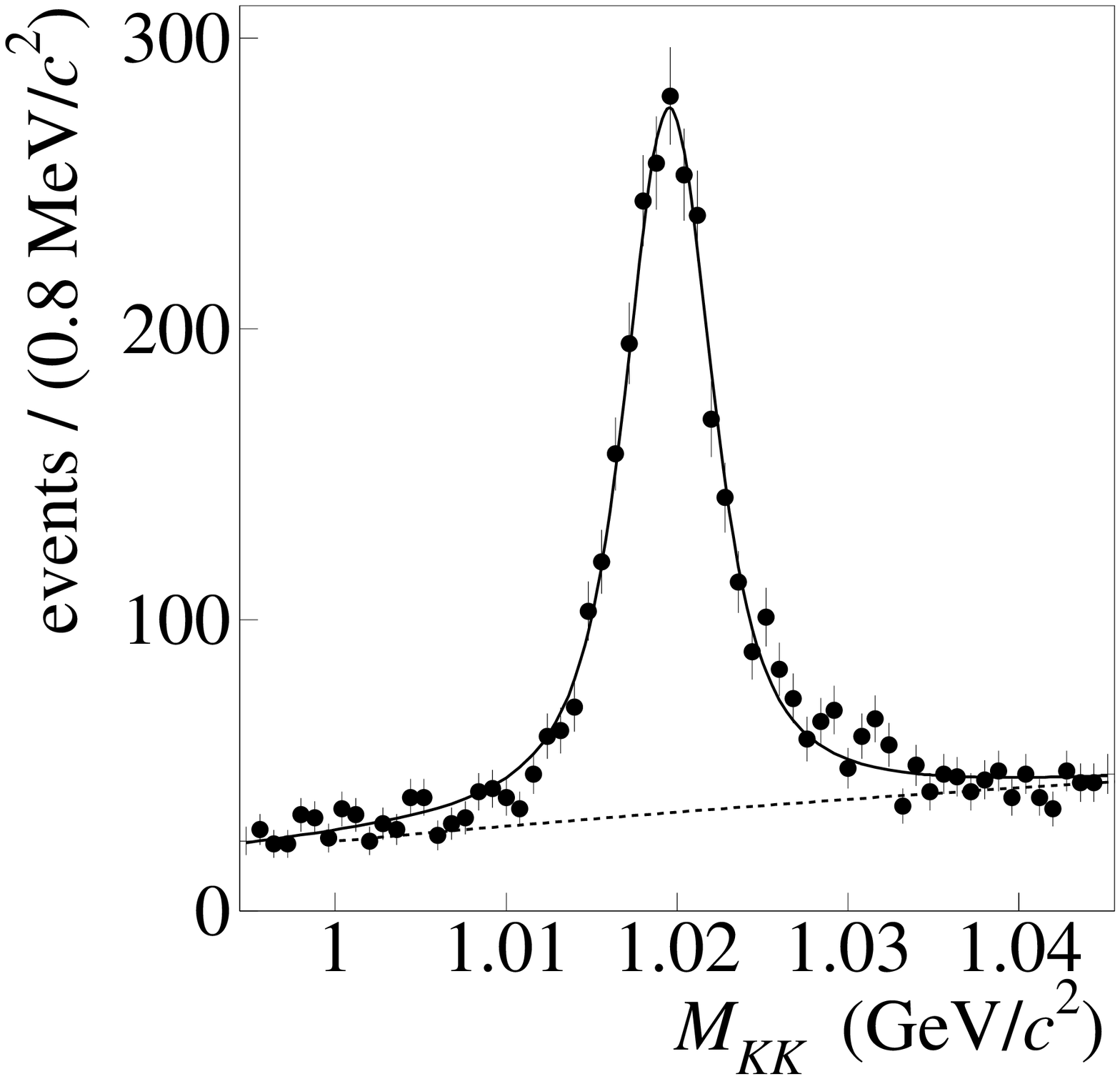,height=4.5cm,width=4.1cm}
            \epsfig{figure=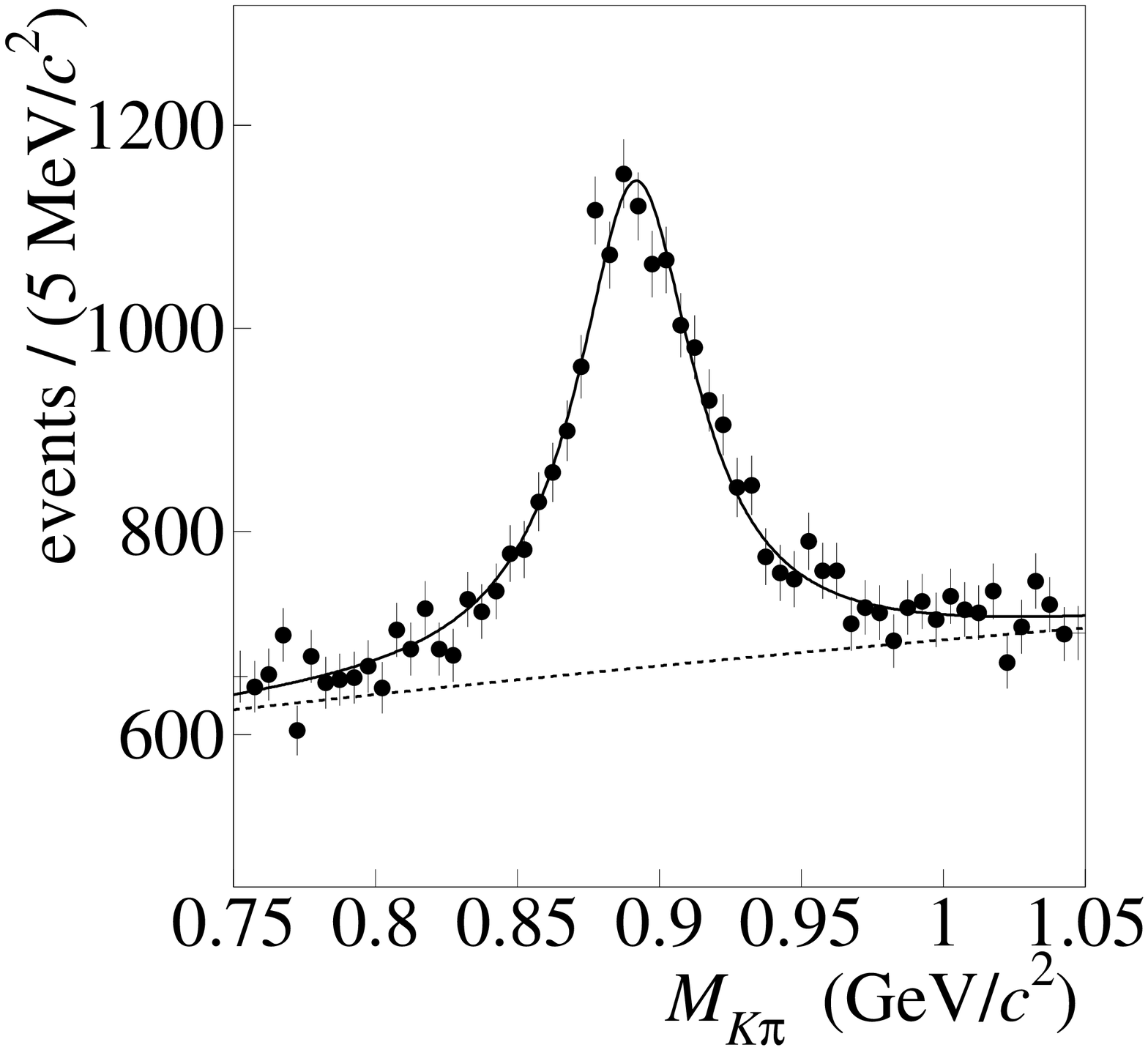,height=4.5cm,width=4.1cm}}
\caption[.]
{The two-kaon invariant mass in the $\phi$ signal region
(left).
Superimposed to the data is the fit to a relativistic $P$-wave
Breit-Wigner for the $\phi$ convoluted with a Gaussian on top 
of a polynomial background.
The mass resolution is 1.1~MeV/$c^2$.
The plot to the right shows a Breit-Wigner fit to the $K^0\pi^+$ invariant mass 
in the $K^{*+}$ signal region.
Both fits use Breit-Wigner parameters from Ref.~\cite{pdg}.
}
\label{fig:phi}
\end{figure}

For $\phi$ candidates, both daughters are required to be kaon
candidates. 
The invariant mass for the $K^+K^-$ pair must lie
within 30~MeV/$c^2$ of the $\phi$ mass
(see Fig.~\ref{fig:phi}).

The natural width of the $K^*$ dominates the resolution in the 
invariant mass spectrum.
For $K^{*0}$ candidates the $K\pi$ invariant mass interval is $\pm$100~MeV/$c^2$.
The selection of $K^{*+}$ comprises $K^+\pi^0$ and $K^0_S\pi^+$ 
combinations within a $K\pi$ mass interval of $\pm$150~MeV/$c^2$
(see Fig.~\ref{fig:phi}).
We require particle identification for the charged daughters 
of the $K^*$.
To suppress combinatorial background we restrict the 
$K^{*+}\rightarrow K^+\pi^0$ helicity angle 
($\cos\theta_H > -0.5$ as defined below). This effectively 
requires the $\pi^0$ momentum to be above 0.35~GeV/$c$.

The helicity angle $\theta_H$ of a $\phi$ or $K^*$ is defined as
the angle between one resonance daughter direction and the 
parent $B$ direction in the resonance rest frame.
For pseudoscalar-vector $B$ decay modes, angular momentum
conservation results in a $\cos^2\theta_H$ distribution,
whereas in decays into vector-vector states, the 
distribution is the result of an {\it a priori\/} unknown superposition of 
transverse and longitudinal polarizations.

We identify $B$ meson candidates kinematically
using two independent variables \cite{babar},
$\Delta E = (E_i E_B -$ \mbox{$\mathbf{p}_i \cdot \mathbf{p}_B - s/2)/\sqrt{s}$} 
and the energy-substituted mass
$m_{\text{ES}} = \sqrt{ (s/2 + \mathbf{p}_i \cdot \mathbf{p}_B)^2 / E_i^2 - 
   \mathbf{p}_B^{\,2} }$,
where $\sqrt{s}$ is the total $e^+e^-$ CM energy.
The initial state four-momentum $(E_i,\mathbf{p}_i)$
derived from the beam kinematics and the four-momentum
$(E_B,\mathbf{p}_B)$ of the reconstructed $B$ candidate 
are all defined in the laboratory.
The calculation of $m_{\text{ES}}$ only involves the
three-momenta of the decay products, and is therefore
independent of the masses assigned to them.
For signal events $\Delta E$ peaks at zero and
$m_{\text{ES}}$ at the $B$ mass. 
Our initial selection requires
$|\Delta E| <$ 0.23 GeV and $m_{\text{ES}} >$ 5.2~GeV/$c^2$.

Monte Carlo (MC) simulation \cite{geant} demonstrates that contamination
from other $B$ decays is negligible.
However, charmless hadronic modes suffer from large backgrounds due to random 
combinations of tracks produced in the quark-antiquark ($q\bar{q}$) 
continuum.
The distinguishing feature of such backgrounds is their 
characteristic event shape resulting from the two-jet production
mechanism.
We have considered a variety of event shape variables in the 
CM that exploit this difference.

One such variable is the angle $\theta_T$ 
between the thrust axis of the $B$ candidate and the thrust axis 
of the rest of the event, where the thrust axis is defined as the 
axis that maximizes the sum of the magnitudes of the longitudinal 
momenta.
This angle is small for continuum events, where the $B$-candidate
daughters tend to lie in the $q\bar{q}$ jets, and uniformly
distributed for true $B\overline{B}$ events.
Thus we require $|\cos\theta_T| < 0.9$ 
(0.8 for $\phi K^{*+}$). 

Other quantities that characterize the event shape are the
$B$ polar angle $\theta_B$ and the angle $\theta_{q\bar{q}}$ 
of the $B$-candidate thrust axis, both defined with respect to the beam axis,
as well as the angular energy flow of the charged particles
and photons relative to the $B$-candidate thrust axis.
For $\Upsilon(4S)$ decays into two pseudoscalar $B$ mesons, 
the $\theta_B$ distribution has a $\sin^2\theta_B$ 
dependence, whereas the jets from continuum events lead to a 
uniform distribution in $\cos\theta_B$.
In $\theta_{q\bar{q}}$, the continuum jets give rise to a 
$(1+\cos^2\theta_{q\bar{q}})$ distribution, 
while the thrust direction of true $B$ decays is random.
We enhance the background suppression by forming an optimized
linear combination of eleven variables (Fisher discriminant):
$|\cos\theta_B|$, $|\cos\theta_{q\bar{q}}|$, and 
energy flow into the nine 10$^\circ$ polar angle intervals coaxial 
around the $B$ candidate thrust axis \cite{CLEO-fisher}.

We use an extended unbinned maximum likelihood (ML) fit to extract
signal yields. 
The extended likelihood for a sample of $N$ events is
\begin{equation}
{\cal L} = \exp\left(-\sum_{i=1}^{M} n_i\right)\, \prod_{j=1}^N 
\left(\sum_{i=1}^M n_i\, {\cal P}_i(\vec{x}_j;\vec{\alpha})\right) ,
\label{eq:likel}
\end{equation}
where ${\cal P}_i(\vec{x}_j;\vec{\alpha})$ describes the probability
for candidate event $j$ to belong to category $i$, based on
its measured variables $\vec{x}_j$, and fixed parameters
$\vec{\alpha}$ that describe the expected distributions of these
variables in each of the $M$ categories.
In the simplest case, the probabilities are summed over two
categories ($M=2$), signal and background.
The decays $B^+\rightarrow\phi K^+$ and $B^+\rightarrow\phi \pi^+$
are fit simultaneously with two signal and two corresponding background categories
($M=4$). 
The event yields $n_i$ in each category are obtained by 
maximizing ${\cal L}$ \cite{minuit}.
Statistical errors correspond to unit changes in the quantity 
$\chi^2 = -2\ln{{\cal L}}$ around its minimum value.
The significance of a signal is defined by the square root of the change in 
$\chi^2$ when constraining the number of signal events to zero in the 
likelihood fit.

The probability ${\cal P}_i(\vec{x}_j;\vec{\alpha})$ 
for a given event $j$ is the product of independent probability
density functions (PDFs) in each of the fit input variables $\vec{x}_j$.
These are $\Delta E$, $m_{\text{ES}}$, 
$M_{KK}$ for all channels, 
$M_{K\pi}$ for the $\phi K^*$ channels, the $\phi$ helicity angle for 
pseudoscalar-vector decays, and event shape quantities as discussed below.
For the simultaneous fit to the decays $B^+\rightarrow\phi K^+$
and $\phi\pi^+$ we include normalized residuals 
derived from the difference between measured and expected 
DIRC Cherenkov angles for the charged 
primary daughter.
Additional separation between the two final states is provided by 
$\Delta E$.

The fixed parameters $\vec{\alpha}$ describing the PDFs are extracted 
from signal and background distributions from MC 
simulation, on-resonance $\Delta E$--$m_{\text{ES}}$ sidebands, and
off-resonance data. 
The MC resolutions are adjusted by comparisons of data and simulation 
in abundant calibration channels with similar kinematics and topology,
such as $B\rightarrow D\pi, D\rho$ with $D\rightarrow K\pi, K\pi\pi$.
The simulation reproduces the event-shape variable distributions
found in data.
The Cherenkov angle residual parameterizations are 
determined from samples of $D^0\rightarrow K^-\pi^+$
originating from $D^*$ decays.  

For the parameterization of the PDFs for $\Delta E$,
$m_{\text{ES}}$, and resonance masses we employ
Gaussian and Breit-Wigner functions to describe the
signal distributions. 
For the background we use low-degree polynomials or,
in the case of $m_{\text{ES}}$, 
an empirical phase-space function \cite{argus}.
The background parameterizations for $M_{KK}$ and $M_{K\pi}$
also include a resonant component to 
account for $\phi$ and $K^*$ production in the continuum.
The $\phi K$ ($\phi\pi$) helicity-angle distribution is
assumed to be $\cos^2\theta_H$ for signal.
The background shape is again separated into contributions 
from combinatorics and from real $\phi$ mesons, both
fit by nearly constant low-degree polynomials.  
The Cherenkov angle residual PDFs are Gaussian
for both the pion and kaon distributions.
The thrust and production angle PDFs are parameterized 
by polynomials, with the exception of the background in 
$|\cos\theta_T|$, where we use an exponential.
The Fisher discriminant is described by an asymmetric Gaussian 
for both signal and background. 

For all modes, we test the fit response
for various choices of preselection and fit strategies
with samples generated according to the PDFs, 
each containing the expected number of events in signal and background. 
Signal yields were found to be unbiased.
In the $\phi K^{0}$ analysis the results of our tests show
that fitting either to $|\cos\theta_T|$ and $\cos\theta_B$ or 
to the Fisher discriminant yields comparable significance.
Thus, we use only the thrust and $B$ polar angle in this analysis.
In the other modes we find that the additional background 
discrimination provided by the Fisher discriminant improves
the expected significances of the results, and use this approach.


\begin{table}
\begin{center}
\caption{Summary of results;
$\varepsilon$ denotes the reconstruction efficiency and
$\varepsilon_{\text{tot}}$ 
the total efficiency including daughter branching fractions, both in percent;
$N$ is the number of events entering the ML fit,
$n_{\text{sig}}$ the fitted number of signal events,
$S$ the statistical significance (in Gaussian $\sigma$), and
${\cal B}$ the measured branching fraction including statistical
and systematic errors. 
The subscripts in the $\phi K^{*+}$ modes refer to the kaon daughter
of the $K^{*+}$.
}
\label{tab:results}
\begin{tabular}{lccccccc}
\hline
\hline
\vspace{-3mm}&&&&&&\\
Mode & $\varepsilon$ & $\varepsilon_{\text{tot}}$  & $N$ 
& $n_{\text{sig}}$ & $S$ & ${\cal B}(10^{-6})$ \cr
\vspace{-3mm}&&&&&&\\
\hline
\vspace{-3mm}&&&&&&\\
$\phi K^+$    & 36.4 & 17.9 & 4202 & $31.4^{+6.7}_{-5.9}$ & 10.5 &   $7.7^{+1.6}_{-1.4}\pm 0.8$  \cr
\vspace{-3mm}&&&&&&\\
$\phi K^0$    & 37.4 & 6.1  &  351 & $10.8^{+4.1}_{-3.3}$ &  6.4 &   $8.1^{+3.1}_{-2.5}\pm 0.8$  \cr
\vspace{-3mm}&&&&&&\\
$\phi K^{*+}$ & --   & 4.9  & --   & --                   &  4.5 &   $9.7^{+4.2}_{-3.4}\pm 1.7$  \cr
\vspace{-3mm}&&&&&&\\
\hline
\vspace{-3mm}&&&&&&\\
~~$\phi K^{*+}_{K^+}$ & 15.1 & 2.5&  781 & $7.1^{+4.3}_{-3.4}$ & 2.7 & $12.8^{+7.7}_{-6.1}\pm 3.2$ \cr
\vspace{-3mm}&&&&&&\\
~~$\phi K^{*+}_{K^0}$ & 21.5 & 2.4&  381 & $4.4^{+2.7}_{-2.0}$ & 3.6 & $ 8.0^{+5.0}_{-3.7}\pm 1.3$ \cr
\vspace{-3mm}&&&&&&\\
\hline
\vspace{-3mm}&&&&&&\\
$\phi K^{*0}$         & 26.3 &  8.6& 2517 & $16.9^{+5.5}_{-4.7}$ & 6.6 & $8.6^{+2.8}_{-2.4}\pm 1.1$ \cr
\vspace{-3mm}&&&&&&\\
$\phi \pi^+$          & 38.9 & 19.1& 4202 & $0.9^{+2.1}_{-0.9}$  & 0.6 & $<1.4$ (90\% CL)      \cr
\vspace{-3mm}&&&&&&\\
\hline
\hline
\end{tabular}
\end{center}
\end{table} 

\begin{figure}[hbt]
\setlength{\epsfxsize}{1.0\linewidth}\leavevmode\epsfbox{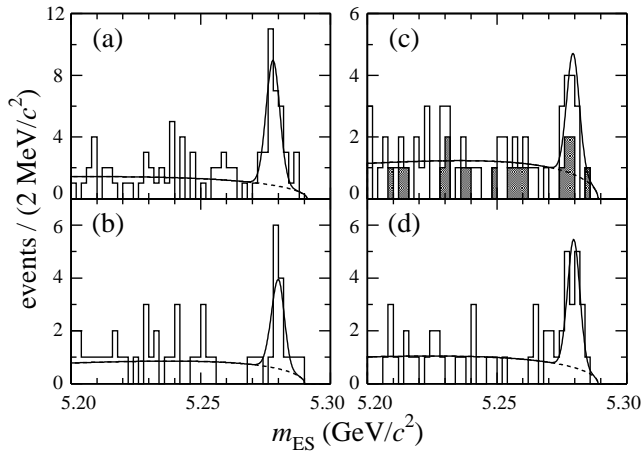}
  \caption{\label{fig:mbproj}%
Projections onto the variable $m_{\text{ES}}$. The histograms show data for
(a) $B^+\rightarrow\phi K^+$;
(b) $B^0\rightarrow\phi K^{0}$; (c) $B^+\rightarrow\phi K^{*+}$; (d)
$B^0\rightarrow\phi K^{*0}$ after a requirement on the signal probability
${\cal P}_{\text{sig}}/\Sigma{\cal P}_i$ with the PDF for $m_{\text{ES}}$
excluded.
In (c) the histogram is the sum of the two $\phi K^{*+}$ channels
while the shaded area is $K^{*+}\rightarrow{K^0\pi^+}$
alone.
The solid (dashed) line shows the PDF projection of the full fit 
(background only).
 }
\end{figure}

The results of our ML fit analyses are summarized in Table~\ref{tab:results}.
For the branching fractions we assume equal production rates of 
$B^0\overline{B}\mbox{}^0$ and $B^+B^-$.
We find significant signals in all four $B\rightarrow\phi K$ and $\phi K^*$
decay modes.
The number of fit events, their statistical significance, and the ML fit $\chi^2$
values are well reproduced with generated samples.
Projections of the input variables are in good agreement 
with the fit results, as shown for 
$m_{\text{ES}}$ in Fig.~\ref{fig:mbproj}.

We check the stability of our results by reducing the number
of input variables in the fit. 
In particular, we find statistically significant signals
even if event shape variables are omitted from the fit and only 
preselection criteria are required.
Correlations among the input variables are found to be less
than 10\%.

Systematic uncertainties in the ML fit originate
from assumptions about the signal and background distributions.
We vary the PDF parameters within their respective uncertainties,
and derive the associated systematic errors.
They range between 4 and 9\% (17\% for the final state that includes a $\pi^0$).
The signals remain statistically significant within 
these variations.

The dominant systematic errors in the efficiency are track 
finding (1.2\% per track), particle identification (2\% per track), 
and $K^0_S$ and $\pi^0$
reconstruction (7\% and 5\%, respectively). 
Other minor systematic effects
from event selection criteria, daughter branching fractions,
MC statistics, and $B$ meson counting sum to less than 4\%.
The efficiency in the ML fit to signal 
samples can be less than 100\% because of fake combinations
passing the selection criteria, and we account for this with a 
systematic uncertainty (2--5\%).
This effect is larger in the $K^*$ final 
states because of broader distributions and
combinatorial $\pi^0$ background.
Uncertainties in the efficiency only affect the branching fraction, 
but not the significance of a result.

In the vector-vector final states  
we average efficiencies for the transverse 
and longitudinal angular polarizations and
assign a systematic error as the rms spread
of a uniform efficiency distribution 
between the two extreme cases 
(6\% in $\phi K^{*+}_{K^0}$, 14\%
in $\phi K^{*+}_{K^+}$, and 2\% in $\phi K^{*0}$).  
We combine the results from the two $K^{*+}$ decay channels using
$\chi^2$ distributions convoluted with the uncorrelated part 
of the systematic errors. 

The fit result for the $B^+\rightarrow\phi\pi^+$ branching fraction is
$(2.1^{+4.9}_{-2.1}\pm 0.5)\times 10^{-7}$.
Given the signal yield of less than one event,
we quote an upper limit obtained by integrating the normalized 
likelihood distribution.
The limit incorporates changes by one standard deviation
from uncertainties in PDFs and the reconstruction efficiency.

Event counting analyses, based on the same variable set $x_j$ 
as used in the fits, serve as cross-checks for the ML fit results. 
The variable ranges are generally chosen to be tighter
in order to optimize the signal-to-background ratio, or upper
limit, for the expected branching fractions. 
We count events in a rectangular signal region in the 
$\Delta E$--$m_{\text{ES}}$ plane, and estimate the background 
from a sideband area.
For $B^+\rightarrow\phi K^+$ we find 43 events in the signal region
(expected background 9.4); the corresponding
numbers are 10 (2.8) for $\phi K^0$, 6 (2.2) for $\phi K^{*+}$, 22 (7.3)
for $\phi K^{*0}$, and 2 (3) for $\phi \pi^+$.
The branching fractions measured using this technique are in 
good agreement with those arising from the ML fit analysis.

In summary, we have observed $B$ decays to $\phi K^+$, 
$\phi K^0$, $\phi K^{*+}$, and $\phi K^{*0}$
with significances, including systematic uncertainties,
of greater than four standard deviations (Table~\ref{tab:results}).
The agreement between the branching fractions of charged and neutral modes
is in accordance with isospin invariance under the assumption
of penguin diagram dominance. 
The decay $B^+ \rightarrow \phi \pi^+$ has both CKM and 
color suppression relative to $\phi K^+$
\cite{theory5}
and is therefore not expected to be observed in the present 
data sample.

We are grateful for the 
extraordinary contributions of our \pep2\ colleagues in
achieving the excellent luminosity and machine conditions
that have made this work possible.
The collaborating institutions wish to thank 
SLAC for its support and the kind hospitality extended to them. 
This work is supported by the
the US Department of Energy
and National Science Foundation, the
Natural Sciences and Engineering Research Council (Canada),
Institute of High Energy Physics (China), the
Commissariat \`a l'Energie Atomique and
Institut National de Physique Nucl\'eaire et de Physique des Particules
(France), the
Bundesministerium f\"ur Bildung und Forschung
(Germany), the
Istituto Nazionale di Fisica Nucleare (Italy),
the Research Council of Norway, the
Ministry of Science and Technology of the Russian Federation, and the
Particle Physics and Astronomy Research Council (United Kingdom). 
Individuals have received support from the Swiss 
National Science Foundation, the A.\ P.\ Sloan Foundation, the Research Corporation,
and the Alexander von Humboldt Foundation.

\vskip12pt 
\hbox{$\ \ \ \ \ \ \ \ \ \ \ \ \ \ \ \ \ \ \ \ \ \ \ \ ${\LARGE{\bf---------------}}}

\vspace{-1cm}

\begin{thebibliography}{99}

\bibitem{theory1}
N.G.~Deshpande and J.~Trampetic,
Phys.\ Rev.\ D {\bf 41}, 895 (1990);
N.G.~Deshpande and X.-G.~He,
Phys.\ Lett.\ B {\bf 336}, 471 (1994);
R.~Fleischer,
Z.\ Phys.\ C {\bf 62}, 81 (1994).

\bibitem{theory3}
I.~Dunietz and J.L.~Rosner,
Phys.\ Rev.\ D {\bf 34}, 1404 (1986).

\bibitem{theory4}
Y.~Grossman and M.P.~Worah,
Phys.\ Lett.\ B {\bf 395}, 241 (1997).

\bibitem{theory5}
R.~Fleischer,
Int.\ J.\ Mod.\ Phys.\ A {\bf 12}, 2459 (1997).

\bibitem{firstphik}
CLEO Collaboration, preprint hep-ex/0101032,
submitted to Phys.\ Rev.\ Lett.

\bibitem{babar}
\babar\ Collaboration, B.~Aubert {\it et al.},
SLAC-PUB-8569, submitted to Nucl.\ Instrum.\ and Methods.

\bibitem{pep} 
PEP-II Conceptual Design Report, SLAC-R-418 (1993).

\bibitem{pdg}
Particle Data Group, D.E.~Groom {\it et al.}, 
Eur.\ Phys.\ J.\ C {\bf 15}, 1 (2000).

\bibitem{CLEO-fisher}
CLEO Collaboration,
D.M.~Asner {\it et al.}, 
Phys.\ Rev.\ D {\bf 53}, 1039 (1996).

\bibitem{geant}
The \babar\ detector Monte Carlo 
simulation is based on GEANT:
R.~Brun {\it et al.}, CERN DD/EE/84-1.

\bibitem{minuit}
F.~James,
CERN Program Library, D506.

\bibitem{argus}
ARGUS Collaboration, H.~Albrecht {\it et al.}, Phys.\ Lett.\ B {\bf 241}, 278 (1990).

\end{thebibliography}
\end{document}